\documentclass[pop,superscriptaddress,showpacs,preprint]{revtex4-1}
\usepackage{bm}
\usepackage{graphicx}
\usepackage{latexsym}
\usepackage{amssymb}
\usepackage{amsmath}
\usepackage{xcolor}
\usepackage{MnSymbol}
\usepackage{wasysym}
\usepackage{color}

\begin{document}

\title{First-Principles Calculation of Principal Hugoniot and K-Shell X-ray Absorption Spectra for Warm Dense KCl }
\author{\surname{Shijun} Zhao}
\author{\surname{Shen} Zhang}
\author{\surname{Wei} Kang}
\email{weikang@pku.edu.cn}
\affiliation{HEDPS, Center for Applied Physics and Technology, Peking University, Beijing 100871, China}
\affiliation{College of Engineering, Peking University, Beijing 100871, China}
\author{\surname{Zi} Li}
\affiliation{Institute of Applied Physics and Computational Mathematics, Beijing 100088, China}
\author{\surname{Ping} Zhang}
\affiliation{HEDPS, Center for Applied Physics and Technology, Peking University, Beijing 100871, China}
\affiliation{Institute of Applied Physics and Computational Mathematics, Beijing 100088, China}
\author{\surname{Xian-Tu} He}
\email{xthe@iapcm.ac.cn}
\affiliation{HEDPS, Center for Applied Physics and Technology, Peking University, Beijing 100871, China}
\affiliation{College of Engineering, Peking University, Beijing 100871, China}
\affiliation{Institute of Applied Physics and Computational Mathematics, Beijing 100088, China}

\begin{abstract}
Principal Hugoniot and K-shell X-ray absorption spectra of warm dense KCl are calculated using the first-principles molecular dynamics method. Evolution of electronic structures as well as the influence of the approximate description of ionization on pressure (caused by the underestimation of the energy gap between conduction bands and valence bands) in the first-principles method are illustrated by the calculation.  Pressure ionization and thermal smearing are shown as the major factors to prevent the deviation of pressure from global accumulation along the Hugoniot. In addition, cancellation between electronic kinetic pressure and virial pressure further reduces the deviation.  The calculation of X-ray absorption spectra shows that the band gap of KCl persists after the pressure ionization of the $3p$ electrons of Cl and K taking place at lower energy, which provides a detailed understanding to the evolution of electronic structures of warm dense matter.
\end{abstract}

\maketitle

\section{INTRODUCTION}

Warm dense matter (WDM) is of particular interest to astrophysics, geophysics, and inertial confinement fusion (ICF).\cite{Nettelmann2012,Lindl2004,Lindl1995} 
Recent developments of high-power laser facilities\cite{Hurricane2014, Zhang2008} and corresponding diagnosing techniques\cite{Barker1972, Miller2007} greatly facilitate its investigation, making WDM accessible to systematical laboratory studies.
Laser-driven shock compression is a typical way to generate warm dense states in large-scale laser facilities, such as Shen-Guang IIU\cite{Zhao2013,Zhang2012}.
Among materials used in shock compression experiments, KCl is a good choice for X-ray absorption measurements\cite{Zhao2013, bradley1987time} considering its large intrinsic band gap of 8.69 eV\cite{Roessler1968}, which efficiently reduces the preheating of diagnosing  X-ray  in the early stages of the experiments\cite{bradley1987time,Zhao2013}. 
As a typical ionic crystalline material (under ambient conditions) composed of heterogeneous species, KCl is also beneficial for illustrating the evolution of electronic structures (e.g., pressure ionization, energy level shifts, and metalization) of heterogeneous ionic materials under warm dense conditions. Those properties can now be readily detected by their X-ray absorption spectra (XAS).
In particular, the variation of the intrinsic band gap under compression has a significant influence on the  degree of ionization, which is a critical parameter in  several commonly used models of equation of state (EOS), e.g., the quotidian equation of state (QEOS) model\cite{More1988}.

From the methodological point of view, KCl is also an illustrating example to display the influence of  the approximate description of ionization, caused by the underestimation of the energy gap between conduction bands and valence bands, in the first-principles molecular dynamics (FPMD) method, now an influential theoretical tool in the study of WDM\cite{Graziani2014}. 
The FPMD method, usually implemented with local density approximation (LDA) or the Perdew-Burke-Ernzerhof (PBE) version of generalized gradient approximation (GGA)\cite{Perdew1996} to the exchange-correlation interaction, has been known to underestimate the energy gap by 40\% - 60\% for nonmetals\cite{Martin2008}, which causes a serious overestimation in the ionization, and was thus postulated to bring about sizable deviations in the calculation of thermal properties of WDM.
However, this conjecture is not well supported by recent studies. The deviation turns out to be small. FPMD calculations on a large variety of materials, including hydrogen,\cite{Beule1999,Holst2008,Lenosky2000} helium,\cite{Kietzmann2007} aluminum,\cite{Mazevet2008,Minakov2014} and iron\cite{Alfe2002}, display a good agreement with measured Hugoniot data. 
These calculations cover a large range of temperature and pressure, at which electrons  in the outermost shell and  inner shell can be ionized. 
How to understand this small effect remains an open problem. The answer to it, however, is important to the improvement of the FPMD method. The ionization of KCl takes place under a condition relatively easy to realize through current WDM experimental facilities, which makes it a favorable example to examine this effect.

Using the FPMD method, we investigate the variation of electronic structures and the influence of the approximate description of ionization to the calculation of pressure in warm dense KCl. The variation of electronic structures is illustrated through the calculation of XAS.
Our results show that the energy gap,  originally between the occupied valence band and unoccupied conduction band under ambient conditions, 
persists after the occurrence of  pressure ionization of the $3p$ electrons of Cl and K taking place at a lower energy.
Pressure ionization and thermal smearing are shown as the major factors to prevent the deviation of pressure caused by the approximate description of ionization from global accumulation along the Hugoniot.
In addition, cancellation between electronic kinetic pressure and virial pressure further reduces the deviation, leading to a small influence of the approximate description of ionization in the FPMD method, in line with the trend illustrated in previous calculations\cite{Beule1999,Holst2008,Lenosky2000,Kietzmann2007,Mazevet2008,Minakov2014,Alfe2002}.

The rest of the article is organized as follows. In Sec. II  methods and numerical details used in the calculation of principal Hugoniots and K-shell XAS are summarized.
In Sec. III, the influence of the approximate description of ionization on the calculation of pressure is discussed through the calculation of principle Hugoniot. In Sec. IV, the variation of electronic structures of KCl in warm dense states is displayed together with the calculation of XAS. We conclude the article with a short summary in Sec. V.

\section{Computational Methods and Numerical Details}
\subsection{Calculation of Principal Hugoniots}
Principal Hugoniot of shock-compressed KCl is calculated through FPMD simulations consisting of 54 atoms, i.e., 27 formula units of KCl.  Finite-temperature density functional theory (FTDFT)\cite{Mermin1965} is employed in the simulation to describe the statistics of electrons. The calculation is performed using the \texttt{Quantum-Espresso} package.\cite{Giannozzi2009a} The  PBE version of GGA\cite{Perdew1996} is used to account for the exchange-correlation interaction. Projector augmented wave (PAW) pseudopotentials\cite{Blochl1994} generated by the \texttt{ATOMPAW} program \cite{Holzwarth2001a} are employed to model the electron-ion interactions. The plane-wave cutoff is set to be 50 Ry, which controls the deviation within 0.01 eV in the calculation of total energy. Only the $\Gamma$ point is used to sample the Brillouin zone in the FPMD simulations. 

For a given density, a series of simulations are carried out at different ionic temperatures, which are controlled by the No\'{s}e-Hoover thermostat\cite{Nose1984}. The electronic states are populated according to the Fermi-Dirac distribution and  thermodynamical equilibrium is maintained by setting the electronic temperature ($T_e$) the same as the ionic temperature ($T_i$). 
In some cases, a $T_e$ different from the value of $T_i$ is used to illustrate the influence of the approximate description of ionization.
Typical dynamic simulations last for 3 $\sim$ 5 ps with an appropriate time step 0.4 $\sim$ 1.0 fs, varying with temperature and density. 
Data presented in the article are collected from the last 2000 steps.

The principal Hugoniot points are determined by solving the Hugoniot equation\cite{Zeldovich2002}
\begin{equation}\label{eq_hugoniot}
   E_1-E_0=\frac{1}{2}(P_1+P_0)(V_0-V_1),
\end{equation}
where $E$ represents the average internal energy, $V$ is the average volume, and $P$ is the pressure. Subscripts 0 and 1 refer to initial and shocked states, respectively. Kinetic pressure of ions, equal to $k_BT/V$, is added manually in the calculation to accurately account for the total pressure. The reference state ($E_0$, $V_0$) is determined from a face-centered cubic (fcc) structure under ambient conditions. $P_0$ is neglected in the calculation as it is several orders smaller than $P_1$. An auxiliary cubic polynomial interpolating function is employed in solving Eq.~(\ref{eq_hugoniot}) to reduce the numerical error less than 1\%.

Note that KCl undergoes a polymorphic phase transition from a fcc lattice (B1, NaCl structure) to a body-centered cubic (bcc) lattice (B2, CsCl structure) during the shock compression \cite{al1963investigation,Kormer1965,Hayes1974,Mashimo2002}. The transition pressure varies from 1.97 to 2.5 GPa, depending on experimental conditions\cite{al1963investigation,Kormer1965,Hayes1974,Mashimo2002}. The calculation of Hugoniot starts from a B2 structure at the pressure around 3.5 GPa, and phase transition points along the Hugoniot are avoided.

\subsection{Calculation of XAS}

The XAS of warm dense KCl is calculated via first-principles methods based on FPMD, similar to those used in previous studies.\cite{Recoules2009, Zhang_arxiv}
Theoretically,  X-ray absorption is characterized by its cross-section $\sigma(\omega)$, which is calculated in a perturbative way  as \cite{Taillefumier2002}
\begin{equation}\label{eq_cross}
\frac{\sigma(\omega)}{4\pi^2\alpha\hbar\omega}=\sum\limits_{f}
\left| \left< f\left| \boldsymbol{\hat{\epsilon}\cdot r} \right| i\right> \right|^2 (1-\mathcal{F}(E_f))
\delta(E_f-E_i-\hbar\omega),
\end{equation}
where, letters $f$ and $i$ represent final and initial states respectively,  $\hat{\epsilon}$ is a unit vector denoting the polarization of incident light, $\alpha$ is the fine structure constant, and $\mathcal{F}(E_f)$ is the Fermi-Dirac distribution of final states at finite temperature.
Calculated using Eq.~(\ref{eq_cross}), K-shell XAS of shock-compressed Al  have displayed a good agreement with experimental measurements\cite{Benuzzi-Mounaix2011, Zhang_arxiv}.
Information on electronic structures and thermodynamical properties of WDM can be further derived through these calculations. \cite{Zhang_arxiv,Benuzzi-Mounaix2011}

The XAS calculation is performed using the \texttt{XSPECTRA} package\cite{Gougoussis2009,Taillefumier2002} supplied with the \texttt{Quantum-Espresso} distribution\cite{Giannozzi2009a}. A minor modification is made to account for the Fermi-Dirac distribution of electrons at finite temperature. Wave functions of the final states used in the XAS calculation are prepared by a density functional theory (DFT) calculation on the atomic configurations generated in the previous Hugoniot calculation.
The plane-wave energy cutoff is 50 Ry, and the hole in the 1$s$ state of the excited Cl atom is manually included in the core of the pseudopotential with the help of the GIPAW pseudopotential\cite{Pickard2001} to approximately account for electron-hole interactions. Other atoms in the system are presented using the Troullier-Martins pseudopotentials\cite{Troullier1991}. The exchange-correlation functional is kept the same as that used in the Hugoniot calculation. With the wave functions prepared, $\sigma(\omega)$ is then calculated through Eq.~(\ref{eq_cross}).
For each thermodynamical condition, a total of 8 XAS  are calculated on selected snapshots along the trajectory of ions, separated by 200 time steps. The final results of XAS are averages of the spectra of each snapshot.

The position of the K-absorption edge, defined as $E_{edge}-E_{1s}$, is calculated in two steps. Firstly, the edge energy with respect to the chemical potential $\mu$,  $E_{edge}-\mu$, is determined directly from the XAS as the intersection of the abscissa to the slope of the edge. Secondly,  the energy of the $1s$ orbital of Cl referring to $\mu$, i.e., $E_{1s}-\mu$, is calculated by an separate FPMD simulation on a system consisting of 16 atoms. 8 atomic configurations separated by 200 time steps are generated by the FPMD simulation. Then, for each configuration, a DFT calculation with pseudopotentials explicitly including all electrons (generated by the \texttt{ATOMPAW} pseudopotential package \cite{Holzwarth2001} with a core cutoff radius of 0.5 Bohr  and  a plane-wave cutoff  energy of 400 Ry) is performed to get $E_{1s}-\mu$.  The difference between these two energies is the K-edge position $E_{edge}-E_{1s}$ required.

{Note that in the second step, the number of ions is remarkably reduced from 54 to 16 because of the extremely high computational costs.  Over 1000 electronic states has to be included to describe a warm dense KCl system consisting of 54 ions at several eV in the DFT calculation. It is challenging to perform such a calculation with all electronic states explicitly described at a plane wave cutoff of 400 Ry. }

To directly compare with the experimentally measured  transition energy, an correction of 92 eV is added to the calculated edge position to compensate the intrinsic inaccuracy of the DFT method in calculating $E_{1s}-\mu$.


\section{Principal Hugoniot and the Influence of the Approximate Description of Ionization}

\begin{figure}[]
   \begin{center}
   \includegraphics[width=0.45\textwidth]{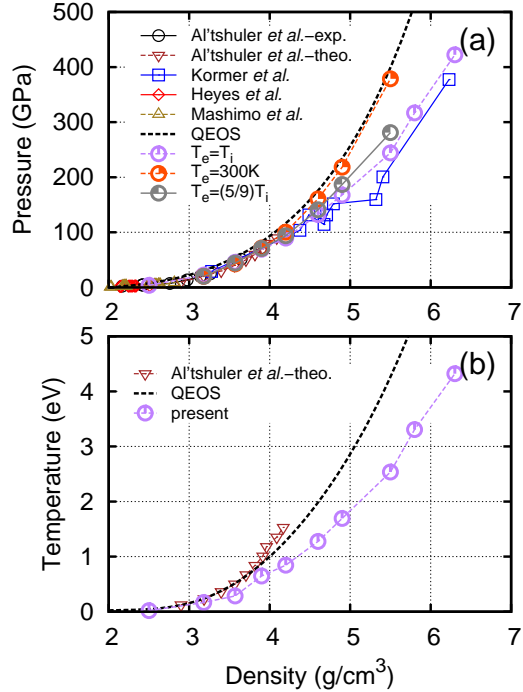}
  \end{center}
   \caption{(Color online) Shocked principal Hugoniots of KCl in (a) P-$\rho$ diagram and (b) T-$\rho$ diagram.  
   }
   \label{hug}
\end{figure}

 Calculated principal Hugoniot of KCl is presented in Fig.~\ref{hug}(a) together with available experimental measurements\cite{Mashimo2002,Kormer1965,Hayes1974}. Theoretical principal Hugoniots derived from the QEOS model\cite{More1988} ({calculated using the \texttt{mpqeos} program included in the \texttt{MULTI} package\cite{Ramis1988}) as well as Altshuler's model\cite{al1963investigation} are also displayed for comparison. Corresponding temperature along the Hugoniot is displayed in Fig.~\ref{hug}(b). Fig.~\ref{hug}(a) shows that the FPMD Hugoniot generally agrees well with experimental data. A small  overestimation about 25 GPa in pressure is observed at densities higher than 5 g/cm$^3$ in comparison with the measurements of Kormer {\it et al.}\cite{Kormer1965} 

The origin of this deviation is not clear so far. On the theoretical side, it is possibly the reflection of the overestimation of ionization in the FPMD method. Calculation on LiF\cite{Clerouin2005} displayed a similar trend compared with Kormer {\it et al.}'s measurements. On the other hand, uncertainties in experimental measurements could also be a cause of the deviation.  Fig.~\ref{hug}(a) shows that the deviation starts from a sharp discontinuity taking place around 5 g/cm$^3$  in the measured Hugoniot. 
This discontinuity does not belong to any of the first-order transitions recognized so far, as have been summarized by Duvall and Graham in Ref.~\onlinecite{Duvall1977}. The corresponding pressure (150 GPa) of the discontinuity is far beyond the pressure of the polymorphic B1-B2 phase transition ($\sim$2 GPa\cite{Mashimo2002}) and the pressure of the solid-liquid phase transition ( 33 GPa $\sim$ 48 GPa).\cite{Duvall1977}

The QEOS Hugoniot lies above the experimental measurements as well as the FPMD Hugoniot. The overestimation is derived from a much higher estimation to the temperature, as displayed in Fig.~\ref{hug}(b).  It is quite unusual that the QEOS has a poor performance on a typical ionic crystalline material in the calculation of Hugoniot. It works reasonably well for other typical ionic crystalline materials like NaCl\cite{More1988} and LiF\cite{Clerouin2005}.
The Altshuler's  model has a similar performance as the QEOS model in the calculation of pressure. However, it gives an even higher estimation to the temperature and tends to overestimate the pressure on further compression. The large variation of temperatures displayed in Fig.~\ref{hug}(b) highlights the necessity to take temperature into account as a critical reference in building theoretical EOS models.

Calculated degree of ionization along the Hugoniot is displayed in Fig.~\ref{gap}(b). Using the FPMD method, the statistics of electron is described by Mermin's functional\cite{Mermin1965}, which assigns an average occupation number to each electronic state according to the Fermi-Dirac distribution.
By this method, the ionization ratio $R(T)$ is estimated as\cite{Kowalski2007}
\begin{equation}\label{ion_r}
R(T)=\frac{1}{N_i\bar{Z}}\int_{\mu}^{\infty}\frac{D(E) dE}{e^{(E-\mu)/k_BT}+1},
\end{equation}
where $N_i$ is the total number of ions, $\bar{Z}$ is the averaged charge number for each ion, $D(E)$ is the the density of state, and $\mu$ is the chemical potential of the system.

As a result of the mass-action law,\cite{Ebeling1968} the ionization ratio $R(T)$ is roughly proportional to $\exp[-(E_g/2k_BT)]$ at modest temperatures, where $E_g$ is the energy of the gap.  For temperature much higher than $E_g/2$, $\mu$ would be much lower than the energy of electrons at the bottom of the shell, as the consequence of finite number of electrons.  Accordingly,  $R(T)$, as estimated by Eq.~(\ref{ion_r}), would approach to 100\%. 
For commonly used LDA and GGA exchange-correlation functionals in the FPMD simulations, $E_g$ is generally underestimated by 40\% - 60\% for nonmetals\cite{Martin2008}, which implies  a sizable overestimation to the $R(T)$ of KCl.

\begin{figure}
\includegraphics[width=0.45\textwidth]{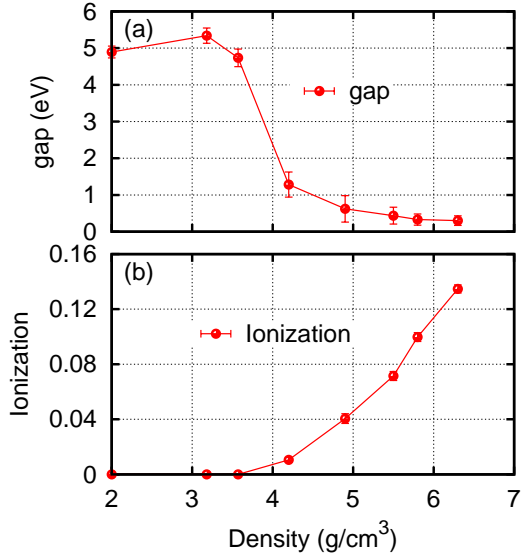}
\caption{\label{gap}(Color online) (a)Energy gap $E_g$ and (b) ionization ratio along the principal Hugoniot calculated using the FPMD simulations. }
\end{figure}

The variation of $E_g$ along the Hugoniot is also an important factor affecting the estimation of $R(T)$. When temperature and density keep increasing along the Hugoniot, $E_g$ approaches to zero as the result of pressure ionization and thermal smearing.
The vanishing of the band gap along the Hugoniot is broadly observed in materials of finite energy gap under ambient conditions, among which LiF\cite{Clerouin2005} is a recent example under extensive studies.
Fig.~\ref{gap}(a) displays the variation of band gap along the Hugoniot, determined through Eq.~(\ref{ion_r})\cite{Kowalski2007}. It shows that the energy gap vanishes around 6.3 g/cm$^3$,  where the $3p$ electrons of Cl are hybridized with the $3d$ electrons of K. This value can also be derived from the variation of  density of states (DOS), as displayed in Fig.~\ref{xanes}(d). 
It provides a threshold indicating when  the system becomes completely metallic and  the inaccuracy caused by the gap becomes negligibly small. Beyond the threshold, the reliability of the FPMD method has been well demonstrated by the calculations on typical  metallic materials, such as Al and Fe\cite{Minakov2014, Alfe2002, Dai2012, Mazevet2008, Zhang_arxiv}. An accuracy of several percentages in both pressure and electronic structures can be achieved in the warm dense region  (before the localized electrons in the lower shell are thermally excited).
These results imply that the inaccuracy of ionization in the FPMD method is constrained by the pressure ionization and thermal smearing, and it will not  accumulate globally, i.e., increase unboundedly, along the Hugoniot.
This is, in indeed, what displays in Fig.~\ref{hug}(a) and in the previous calculations\cite{Beule1999,Holst2008,Lenosky2000,Kietzmann2007,Mazevet2008,Minakov2014,Alfe2002}. Extending this observation to a much higher temperature region (above 100 eV) strongly depends on the temperature effect of exchange-correlation functionals. Recent progress on this topic can be found elsewhere\cite{Graziani2014,Sjostrom2014}.

\begin{figure}
   \begin{center}
   \includegraphics[width=0.45\textwidth]{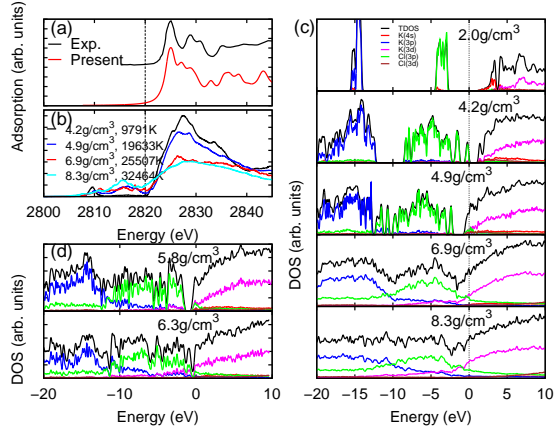}
  \end{center}
   \caption{(Color online) K-XAS of Cl and corresponding PDOS for KCl. (a) XAS calculated under ambient conditions; and (b) XAS under typical warm dense conditions. (c) Corresponding PDOS for (a) and (b), where an energy shift of 2820 eV is applied to move the chemical potential to the vicinity of the origin point. (d) PDOS before and after the vanishing of energy gap .
   }
   \label{xanes}
\end{figure}

\begin{table*}[]
\caption {Contributions of ionic kinetic pressure $P_{i,k}$, electronic kinetic pressure $P_{e,k}$, and virial pressure $P_{pot}$ to  the total pressure (in GPa) at selected electronic temperature and ionic temperature for thermal states $\rho$ = 3.9 g/cm$^3$ and 4.9 g/cm$^3$ corresponding to typical ionization states along the principal Hugoniot.
}
\centering
\begin{tabular} {c c c c c c c c c c c c c}
\hline
\hline
&    & &             &             3.9 g/cm$^3$     &  &&& 4.9 g/cm$^3$    & \\
\hline
&  &  & P$_{i,k}$ &  P$_{e,k}$  & P$_{pot}$ &total &P$_{i,k}$ &P$_{e,k}$ &P$_{pot}$ & total    \\
\hline
&  &  T$_e$=T$_i$              &5.07&1262.82&-1204.83&63.06   &      &     &   & \\
&T$_i$=0.5eV & T$_e$=(5/9)T$_i$&5.04&1268.42&-1205.98&67.47   &      &     &   & \\
&           &  T$_e$=300K      &5.06&1267.07&-1205.67&66.46   &      &     &   & \\
\hline
& &  T$_e$=T$_i$                &10.10&1281.90&-1210.36&81.64  &      &    &   & \\
& T$_i$=1.0eV &T$_e$=(5/9)T$_i$ &10.06&1279.66&-1209.00&80.72  &      &    &   & \\
&           &  T$_e$=300K       &10.10&1280.92&-1208.90&82.12  &      &    &   & \\
\hline
&  &  T$_e$=T$_i$               &20.19&1305.98&-1218.50&107.67  &25.37 & 1699.98  &-1547.22 &  178.13\\
& T$_i$=2.0eV & T$_e$=(5/9)T$_i$&20.20&1299.31&-1214.45&105.06  &25.51 & 1694.25  &-1541.46 &  178.29\\
&           &  T$_e$=300K       &20.22&1296.38&-1212.94&103.66  &25.39 & 1691.45  &-1538.74 &  178.10\\
\hline
&  &  T$_e$=T$_i$                 &  &  &  &  &38.02 & 1735.11  &-1563.49 &  209.63 \\
& T$_i$=3.0eV &  T$_e$=(5/9)T$_i$ &  &  &  &  &38.09 & 1714.11  &-1550.48 &  201.72 \\
&           &  T$_e$=300K         &  &  &  &  &37.98 & 1705.38  &-1543.42 &  199.94 \\
\hline
&  &  T$_e$=T$_i$                   & & & & &50.81 & 1779.85  &-1582.51 &  248.16 \\
&  T$_i$=4.0eV  &  T$_e$=(5/9)T$_i$ & & & & &50.78 & 1735.46  &-1558.39 &  227.86 \\
&      &  T$_e$=300K                & & & & &50.72 & 1722.14  &-1547.44 &  225.42 \\
\hline
\hline
\end{tabular}
\label{tab-pre}
\end{table*}

To quantitatively estimate how an approximate description of ionization in FPMD affects the Hugoniot result, we manually adjust the temperature of electrons. Besides the Hugoniot calculated at $T_e$ = $T_i$, Fig.~\ref{hug}(a) also displays the Hugoniots of $T_e$ = 300 K,  and $T_e$ = (5/9)$T_i$. The fraction of 5/9 gives an approximate correction to the overestimation of ionization caused by the underestimated band gap
for the $\rho <$ 4.0 g/cm$^3$ ($T_e<E_g/2$) part in the Hugoniot curve.  The calculation at $T_e=T_i$ gives a better estimation to the pressure for the $\rho > $ 6.3 g/cm$^3$ part, when the energy gap vanishes and the material is completely metallic. Using these two calculations as references, the deviation of the calculation can be estimated as the smallest difference from the $T_e$ = $T_i$ or $T_e$ = (5/9)$T_i$ curves to the experimental measurements. As displayed in Fig.~\ref{hug}, the largest deviation in the calculation takes place in the region between $\rho$ = 4 g/cm$^3$ and $\rho$ = 6 g/cm$^3$. Below that, the $T_e$ = 5/9 $T_i$ curve only has a small correction of 1 $\sim$ 2 GPa to the $T_e$ = $T_i$ curve because of the small magnitude of $R(T)$, as displayed in Fig.~\ref{gap}(b). The largest deviation is about 50 GPa ($\sim$ 25 \%) at $\rho \sim$ 5.3g/cm$^3$, including experimental uncertainties. This gives an estimation to the worst scenario in the Hugoniot calculation.

To clarify the underlying mechanism of the small influence of the approximate ionization, a detailed decomposition of pressure is given in Table~\ref{tab-pre} for selected electronic and ionic temperatures, corresponding to Hugoniot points at $\rho$ = 3.9 and 4.9g/cm$^3$. As displayed in Fig.~\ref{gap}(a), they represent the starting and ending points of the sharp transition of the band gap along the Hugoniot. 
In Table~\ref{tab-pre}, the pressure is divided into three parts according their origins as
$
P(T,\rho)=P_{e,k}(T,\rho)+P_{i,k}(T,\rho)+P_{pot}(T,\rho).
$
Subscripts $e,k$ and $i,k$ represent kinetic pressures of electrons and ions respectively, and subscript $pot$ refers to virial pressures contributed by  ion-ion, electron-electron, and ion-electron interactions. According to the virial theorem \cite{Martin2008}, $P_{k}$ is  proportional to $2E_k$, and $P_{pot}$ is approximately proportional to $E_{pot}$. 

Table~\ref{tab-pre} displays the change in $P_{pot}$ and $P_{e,k}$ brought about by different ionizations. Generally,  $P_{e,k}$ grows with the increase of ionization, whereas $P_{pot}$ goes to the opposite direction. The fluctuation at low $T_e$ ($T_e <$ 0.6 eV in the table) is caused by electronic excitations from localized electronic states to non-localized states near chemical potential. This is a characteristic of ionic materials. For KCl, the fluctuation is caused by the excitation of the localized $3p$ electrons of Cl to the non-localized $4s$ and $3d$ electrons of K. Most of the changes are brought by the variation of $P_{e,k}$. The direction of the change depends on the competition between the average kinetic energies of the localized and non-localized electrons. 

The cancellation between the increase of $P_{e,k}$ and decrease of $P_{pot}$ leads to a much smaller change in the total pressure. For the Hugoniot point at $\rho$ = 3.9 g/cm$^3$ ( $T_i\sim$ 1.0 eV), the $T_e = (5/9) T_i$ calculation gives a better estimation to the ionization. Its correction to the total pressure is less than 1 GPa, which is about 1\% of the FPMD result. For the Hugoniot point at $\rho$ = 4.9 g/cm$^3$ ( $T_i\sim$ 2.0 eV) the correction calculated with $T_e=(5/9)T_i$ is less than 0.5 GPa. The real correction should be much smaller than that, because the energy gap is now close to zero, as displayed in Fig.~\ref{gap} (a).

\section{ K-shell XAS of C\lowercase{l} and Evolution of Electronic Structures}
 XAS provide information on both ionization and electronic structures. Owing to the thermal excitation of electrons in WDM, electronic structures 10-20 eV below the chemical potential can also be detected by XAS. This is quite different from the low temperature cases, where only electronic structures above or close to the chemical potential is available for X-ray transitions. 
Accordingly, terminologies commonly used in XAS of condensed matter should be used cautiously to avoid discrepancies. 
Here, we do not distinguish sub-categories of XAS and  focus on absorption properties near the beginning of K-absorption spectra ( usually called the edge region) for Cl ions in the shock-generated warm dense KCl. 
Note that, the absorption edge in WDM could be far below (or above in some extreme cases) the chemical potential, as will be illustrated in the following part of this section.

The K-absorption spectra under ambient conditions are compared with experimental measurements\cite{Trischka1945} in Fig.~\ref{xanes}(a). It serves as a reference illustrating how the calculated spectra can be interpreted. Fig.~\ref{xanes}(a) shows that the calculation can reproduce the general feature of the spectra with small difference in details. The energy gap issue is paid special attention in the calculation. A correction of 3.7 eV obtained from a separate G$_0$W$_0$ calculation\cite{hybertsen1986} is added to the FPMD energy gap ($\sim$ 5 eV), so that the calculated gap value agrees with the experimental measurement of 8.69 eV \cite{Roessler1968}.  Hybrid functional HSE06 \cite{Heyd2003} only slightly improves the gap to about 6 eV, and therefore is not helpful to the gap issue. 
Note that the G$_0$W$_0$ correction is not included in the XAS calculation of warm dense states, because the accuracy of the method relies on the description of ionization in the FPMD method, which is generally overestimated  before the vanishing of the gap. Some kind of self-consistent GW approach \cite{Schilfgaarde2006} is necessary to get a theoretically consistent result.

\begin{figure}
   \begin{center}
   \includegraphics[width=0.4\textwidth]{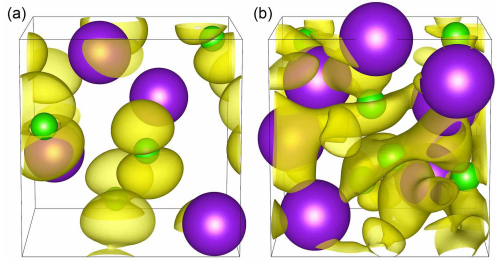}
  \end{center}
   \caption{(Color online) Probability distribution of the $3p$ wave functions of KCl at $E=$ 2816 eV.  (a) The $3p$ wave functions of Cl under ambient conditions. (b) The hybrid $3p$ wave functions of Cl and K at 5.8 g/cm$^3$. The K and Cl ions are represented by purple and green balls, respectively. The isovalue is set to be 0.001.
   }
   \label{orbitals}
\end{figure}

Fig.~\ref{xanes}(b) displays calculated K-absorption spectra of warm dense KCl. 
The prominent feature of the spectra is the persistence of the energy gap (around 2820 eV) under compression, disappearing at a high density $\sim$ 6.3 g/cm$^3$.  The figure also shows that the position of the chemical potential, around 2820 eV, is insensitive to the variation of density and temperature, which is similar to what was found in warm dense Al\cite{Zhang_arxiv}. 

Projected density of states (PDOS) in Fig.~\ref{xanes}(c) and (d) show that the electronic states across the gap have different origins. The majority of the electronic states below the gap consist of the $3p$ states of Cl and K; whereas those above the gap mainly come from the $4s$ and $3d$ states of K. 
The figures also show a strong hybridization between the $3p$ orbitals of Cl and K under compression. This trend is further illustrated  in Fig.~\ref{xanes}(b) by the increase of  band width in the sub-band below the gap (below 2820 eV).  According to the theory of pressure ionization\cite{saltzman_book, Kremp2005}, this hybridization of electronic states of different ions is an evidence of the pressure ionization, i.e., a transition from localized states to non-localized states. This transition is also confirmed by the variation of wave functions of the $3p$ orbitals at  $E$ = 2816 eV, as displayed in  Fig.~\ref{orbitals} for warm dense states $\rho$ = 2.0  g/cm$^3$ ($T=$ 300K) and $\rho$ = 5.8 g/cm$^3$ ($T$ = 3.3 eV). Here, 2816 eV is the energy of the maximum PDOS for the $3p$ states of Cl under ambient conditions, and the thermal state of $\rho$ = 5.8 g/cm$^3$ is the last state in the calculation just before the closing of the gap. The PDOS after the closing of gap is also displayed in Fig.~\ref{xanes}(d). These results show that the $3p$ states of both Cl and K are part of the continuum (non-localized states), and the gap around $E$ = 2820 eV is in the middle of unbounded states above the bottom of the continuum. 

\begin{figure}
   \begin{center}
   \includegraphics[width=0.4\textwidth]{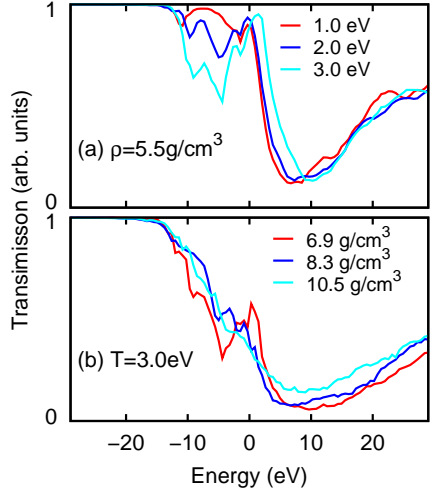}
  \end{center}
   \caption{(Color online) Transmission spectra of warm dense KCl (a) at fixed density $\rho=8.3$ g/cm$^3$; and (b) at fixed temperature $T=3.0$ eV.
  }
   \label{TV}
\end{figure}

Besides the absorption cross section $\sigma$, the transmission (or attenuation) of the material\cite{Zhao2013, bradley1987time} was also measured in experiments. On many occasions, it was used to determine the absorption edge positions\cite{bradley1987time, Zhao2013}. The transmission is proportional to $\exp(-\sigma \tilde{h})$, where $\tilde{h}$ is the areal number density. Fig.~\ref{TV} displays the dependency of transmission spectra (corresponding to the K-absorption of Cl) on density and temperature. The $\tilde{h}$ here takes the value of 790 \AA$^{-2}$,  derived from recent experiment of Zhao {\it it al.} \cite{Zhao2013}  Fig.~\ref{TV}(a) displays the variation of transmission as a function of temperature, at a fixed density of 8.3 g/cm$^3$. It shows that the absorption edge decreases with the growth of temperature, which is mainly caused by the thermal broadening of depopulated electronic states below the chemical potential. 
Fig.~\ref{TV}(b) presents the transmission spectra at fixed temperature of 3 eV, showing that the absorption edge is insensitive to the variation of density. Similar trend has also been found in warm dense Al,\cite{Zhang_arxiv} indicating that the K-absorption edge is useful to diagnose the temperature of warm dense state for a variety of materials.
The feature corresponding to the energy gap appears as a peak  around 2820 eV  in Fig.~\ref{TV}, which gives an explanation to the ``bump'' feature observed in the time-resolved transmission spectra measured by Bradley {\it et al.} \cite{bradley1987time}

\begin{figure}
   \begin{center}
   \includegraphics[width=0.4\textwidth]{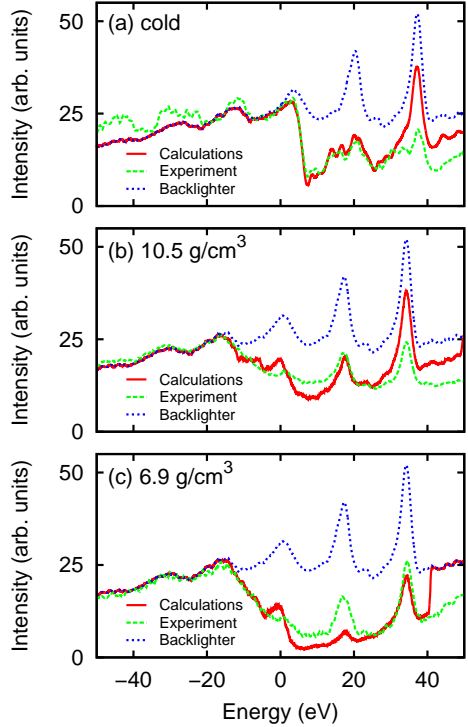}
  \end{center}
   \caption{(Color online) Transmitted X-ray Intensity of KCl for (a) cold samples; (b) 10.5 g/cm$^3$; and (c) 8.3 g/cm$^3$. Experimental measurements and intensity of backlight are also presented for comparison. The energy scale is shifted by 2820 eV to put the chemical potential at the vicinity of the origin point.
   }
   \label{compare}
\end{figure}

To compare with the experiments of Zhao {\it et al.}, \cite{Zhao2013} in which the intensity of transmitted X-ray was measured, the intensity $I(E)$ is further derived from $\sigma$ according to 
$
I(E)=I_0(E)\exp[-\sigma(E)\tilde{h}],
$
with $I_0(E)$ the intensity of backlighter.
Fig.~\ref{compare} supplies a comparison between calculated and measured $I(E)$. The background is deducted according to the left part of the measured spectra below the absorption edge. The intensity of backlighter\cite{Zhao2013} measured in a separate experiment is also included as references, and the energy scale is shifted by 2820 eV so that the position of the chemical potential is located at the vicinity of the origin point.
Fig.~\ref{compare}(a) displays the calculation of KCl under ambient conditions. It well reproduces the XAS up to $\sim$30 eV above the absorption edge. 
Fig.~\ref{compare}(b) and (c) show that our calculation is in reasonable agreement with the experimental measurements up to the chemical potential. For the next 20 eV above that, the calculated intensity is lower than the experimental measurements by about 50\%, which suggests that a refined theoretical treatment is necessary to deal with the spectra above the energy gap in warm dense materials.

\section{SUMMARY}

In summary,  we investigate the variation of electronic structures and the influence of the approximate description of ionization to the calculation of pressure through first-principles calculation of principal Hugoniot and XAS for warm dense KCl. We show that pressure ionization and thermal smearing are the major factors to limit the deviation in the calculation of pressure. They prevent global accumulation of such deviations along the Hugoniot. In addition, the cancellation between kinetic pressure of electrons and virial pressure of interactions further reduces the error.  The calculation of XAS shows that the band gap of KCl persists after the occurrence of the pressure ionization of the $3p$ electrons of Cl and K, which provide a detailed understanding to the evolution of electronic structures for WDM.

\begin{acknowledgments}
We appreciate useful discussions with Jiamin Yang and Yang Zhao on the details of experiments. This work is financially supported by the NSFC (Grant No.  11274019).
\end{acknowledgments}

\bibliographystyle{apsrev4-1}
\bibliography{XANES}

\end{document}